\documentclass{epl}

\usepackage{graphicx}

\def\jcp#1#2#3{{J. Chem.  Phys.} {\bf #1}, (#2) #3}

\def\jpa#1#2#3{{J. Phys. A: Math. Gen.} {\bf #1} (#2), #3}

\def\pra#1#2#3{{Phys. Rev. A.} {\bf #1}, (#2) #3}
\def\prb#1#2#3{{Phys. Rev. B.} {\bf #1}, (#2) #3}
\def\pre#1#2#3{{Phys. Rev. E.} {\bf #1}, (#2) #3}
\def\prl#1#2#3{{Phys. Rev. Lett.} {\bf #1}, (#2) #3}

\def\prs#1#2#3{{Proc. R. Soc.} {\bf #1}, (#2) #3}

\title{\bf On the orientational ordering of long  rods on a lattice}

\author{Anandamohan Ghosh\footnote{ananda@theory.tifr.res.in}
and Deepak Dhar\footnote{ddhar@theory.tifr.res.in} }

\institute{ 
Department of Theoretical Physics, 
Tata Institute of Fundamental Research, \\
Homi Bhabha Road, Mumbai 400 005, India.
}


\begin{document}

\maketitle

\begin{abstract}
We argue that a system of straight rigid rods of length $k$ on square
lattice with only hard-core interactions shows two phase transitions as a
function of density $\rho$ for $ k \geq 7$. The system undergoes a phase
transition from the low-density disordered phase to a nematic phase as
$\rho$ is increased from $0$ at $\rho = \rho_{c1}$, and then again 
undergoes
a reentrant  phase transition from the nematic phase to a disordered
phase at $\rho=\rho_{c2}< 1$.
 \end{abstract}

\section{Introduction} \label{sec:intro}

The study of systems of long rod-like molecules in solution with only
excluded volume interaction has a long history. Onsager had noted that
such a system would show long-range orientational order at large enough
density \cite{Ons49}. Flory studied a lattice model of long rod-like
molecules, and based on a mean-field approximation, argued that the
lattice model would also show an isotropic-nematic phase transition as a
function of density \cite{Flo56}. Zwanzig studied a model of hard rods in
the continuum, but with only a finite number of allowed orientations
\cite{Zwa63}. For the continuum problem, there is general agreement that 
in
three dimensions, the isotropic-nematic transition occurs at large enough
density, if the length-to-width ratio of the molecules is large enough
\cite{Shu04}. In the case of two-dimensional systems of hard needles in 
a plane, there is no
spontaneous breaking of continuous symmetry, but the high-density phase
has orientational correlations that decay as a power-law \cite{Fre85, Kha05}.

The situation is much less clear in the case of lattice models of straight
hard rods of length $k$. The only analytically soluble case is that of
dimers ($ k=2$), for which it is known that the orientational correlations
have an exponential decay for all densities, except the full-coverage limit,
which shows power-law correlations in all dimensions \cite{Hei72,Hus03}.
Power-law correlations at $\rho = 1$ were
also seen in numerical study  on the square
lattice for $k =3$\cite{Gho06}, and $k =4$\cite{Jacobson}.  
Early Monte Carlo studies of
semi-flexible $k$-mers on a square lattice showed no long-range
orientational order for any density or temperature \cite{Bau84}.  
DeGennes and Prost, in 1995, in their book on liquid crystals, wrote ``it
is not quite certain that such a lattice model, even for large $p$ (long 
molecule), will lead to a transition''
\cite{Degennes}.  While there has been some work on this general  problem since
then, the question still remains unsettled \cite{refs} .

In this paper, we present the results of our Monte Carlo study of this
model on the square lattice. We find very strong numerical evidence that
the system shows nematic order at intermediate densities for $k \geq 7$.  
We are not able to study systems near $\rho =1$ in our simulations,
because the relaxation time increases very fast with density. However,
there is fairly convincing numerical evidence (but no proof), that there
is no long-range order at $\rho =1$, and so there must be a second phase
transition as density is increased. To the best of our knowledge, this
second phase-transition has not been discussed in published literature
before. We estimate the critical density for the second transition. The
different phases are shown schematically in Fig. 1.

\begin{figure} \includegraphics[scale=.4, angle=0]{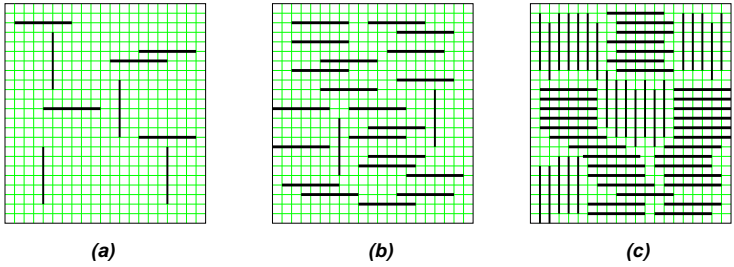} 
\caption{Schematic representation of different phases of the hard rod 
system on a square lattice. (a) The low-density disordered phase (b) 
Intermediate density ordered phase (c) the high density disordered phase. 
} \label{fig:fig1} \end{figure}


\section{Model} \label{sec:model}

Our model consists of a system of hard straight rods, each of which
occupies $k$ consecutive sites on a  lattice.  A rod
has only two possible orientations, horizontal or vertical. We will call
each such rod a $k$-mer.  The only interaction between different rods is
hard core exclusion: no site can be occupied by more than one $k$-mer.

   We take the lattice to be a two-dimensional square lattice of size  $L 
\times L $.   Let $N(n_h,n_v)$ be the number of distinct configurations with 
$n_h$ horizontal and $n_v$ vertical $k$-mers. We define $z_h$ and $z_v$ as 
the activities for the horizontal and vertical $k$-mers, and define the 
grand partition function

\begin{equation}
\Omega( z_h, z_v) = \sum_{n_h,n_v} N(n_h, n_v) z_h^{n_h} z_v^{n_v}.
\end{equation}

The average values of $n_h$ and $n_v$ in the grand canonical ensemble are
obtained by taking partial derivatives of $\Omega$ with respect to $z_h$
and $z_v$. We define the mean density $\rho$ as the fractional number of
sites occupied by $k$-mers. The nematic order parameter $Q$ is defined as

\begin{eqnarray} 
Q = Lim_{ z_v \downarrow z_h} Lim_{L \rightarrow \infty}
\frac{\langle n_v - n_h \rangle}{\langle n_v + n_h \rangle},
\label{eq:ord} 
\end{eqnarray} 
where the limit is taken with $z_v$ tending
to $z_h$ from above. We will usually be interested in the case $z_v =
z_h$, and denote the common value by $z$ without a subscript.

\section{Results of the Numerical Simulations} \label{sec:INT}

\begin{figure}
  \begin{center}
    \begin{tabular}{ccc}
    \resizebox{45mm}{!}{\includegraphics[height=45mm, width=45mm]{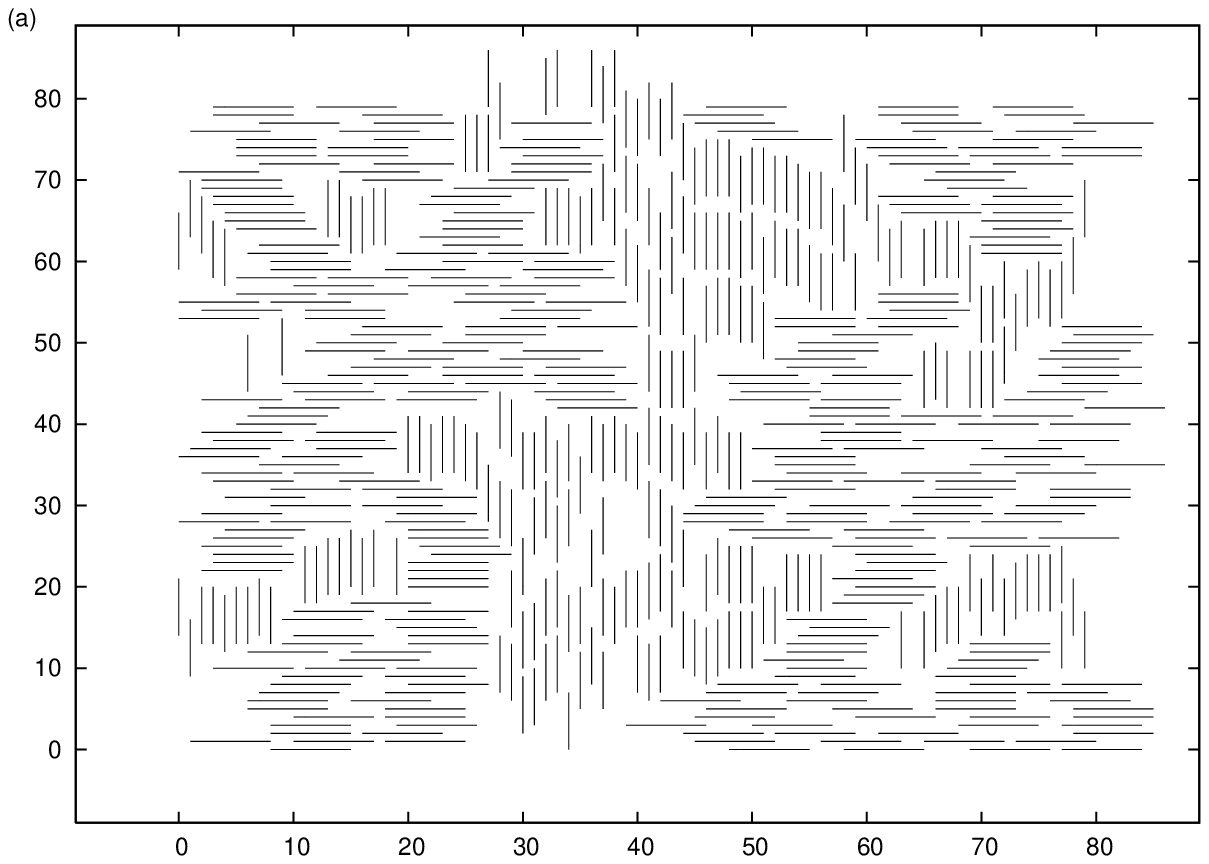}} &
    \resizebox{45mm}{!}{\includegraphics[height=45mm, width=45mm]{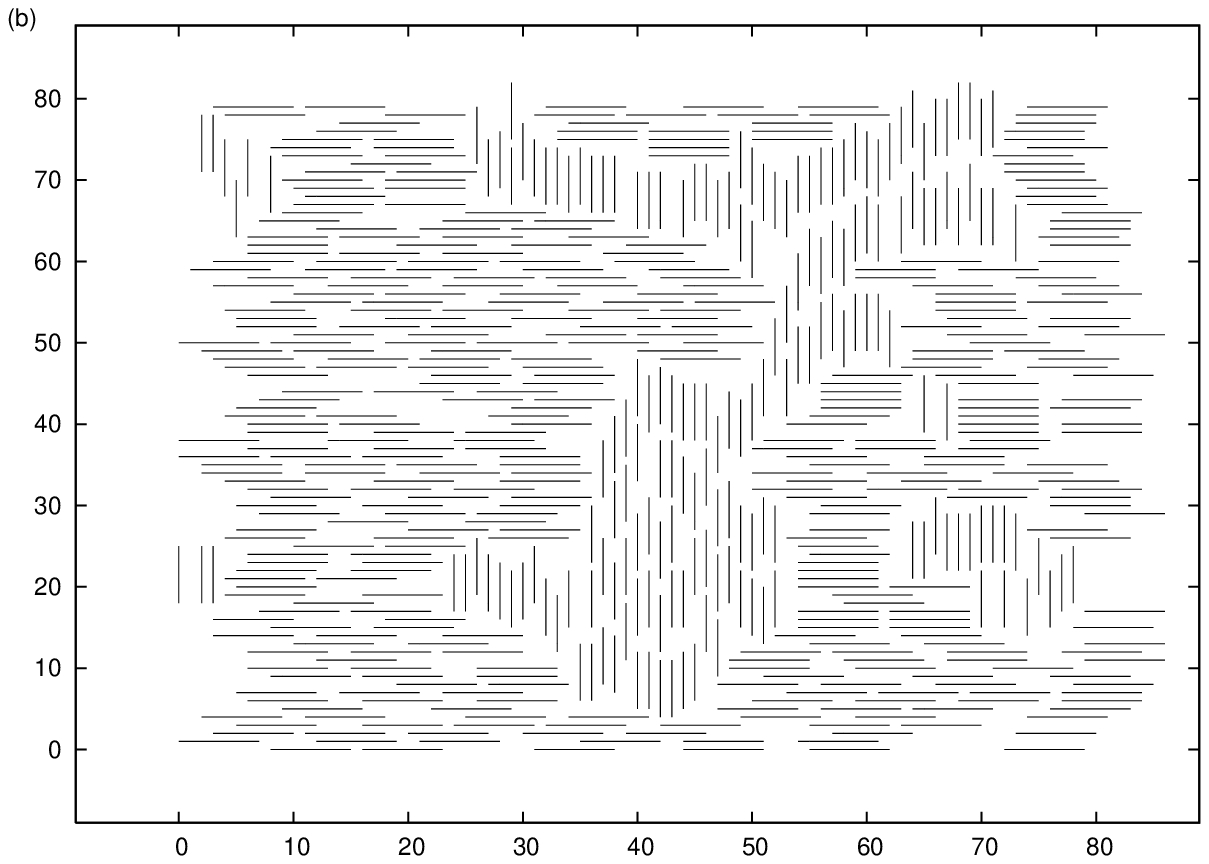}} &
    \resizebox{45mm}{!}{\includegraphics[height=45mm, width=45mm]{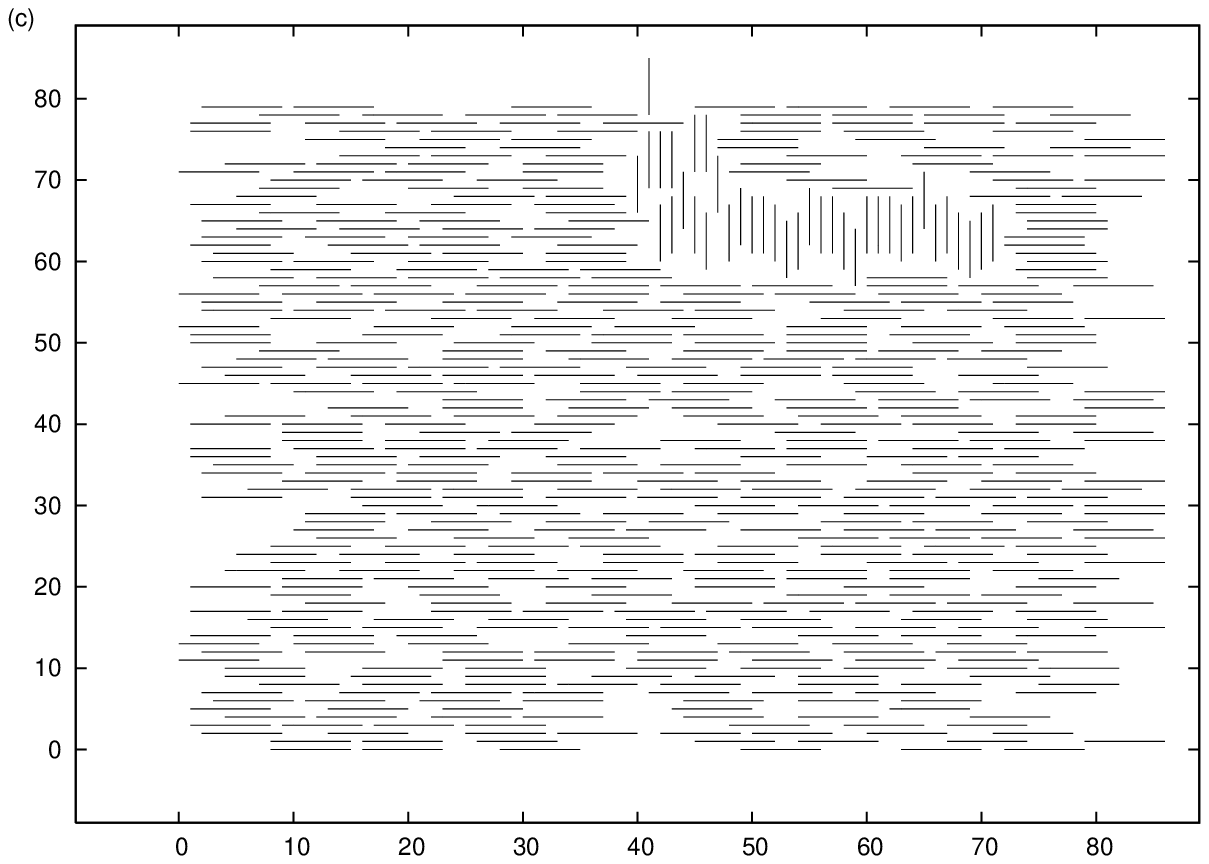}} \\
    \resizebox{45mm}{!}{\includegraphics[height=45mm, width=45mm]{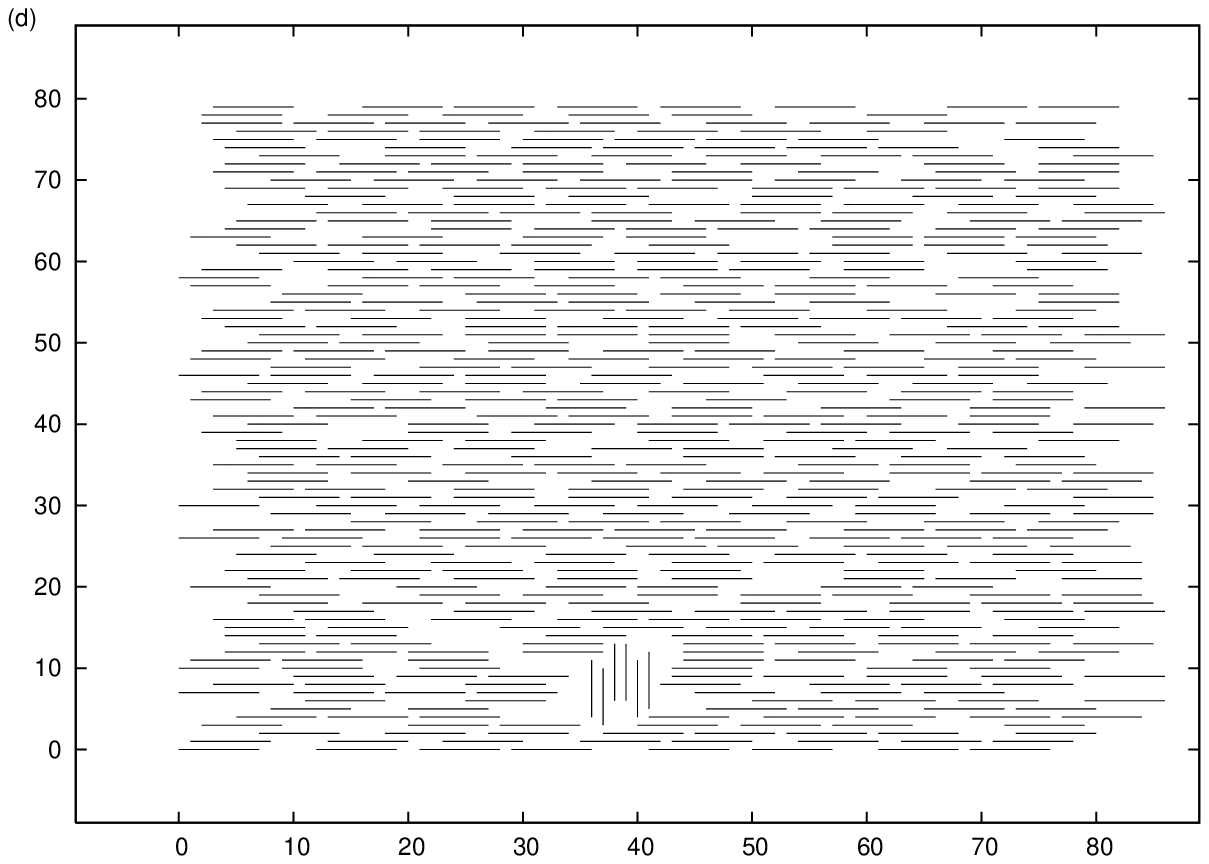}} &
    \resizebox{45mm}{!}{\includegraphics[height=45mm, width=45mm]{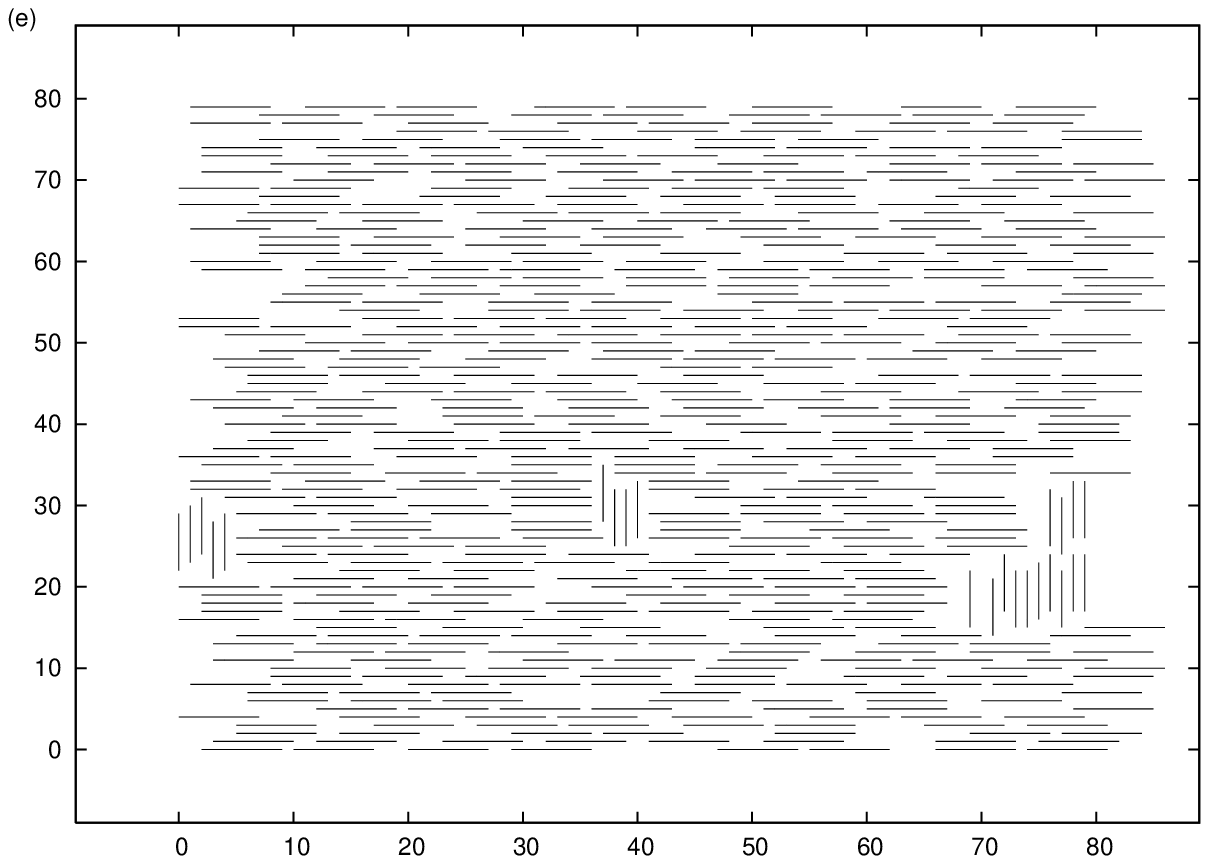}} &
    \resizebox{45mm}{!}{\includegraphics[height=45mm, width=45mm]{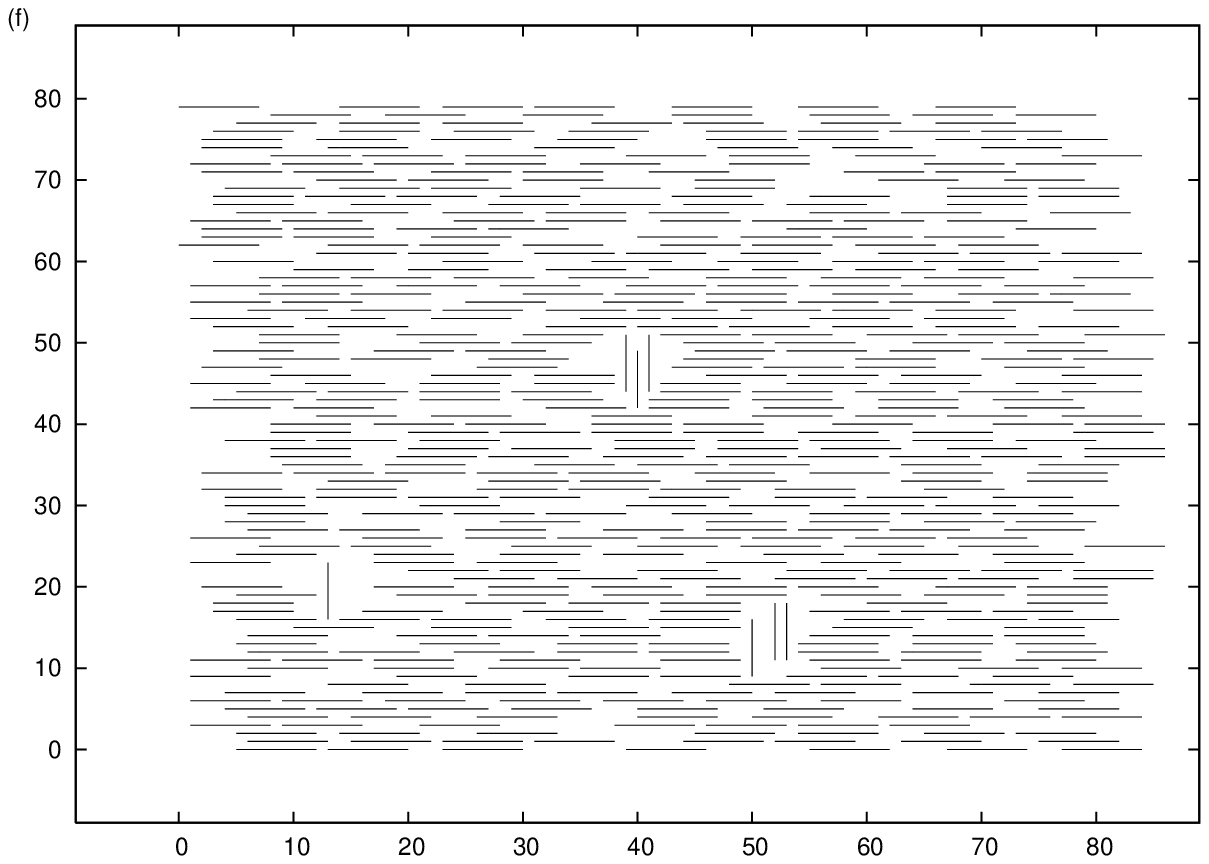}} \\
    \end{tabular}
    \caption{Snapshots of time evolution of $k$-mers, $k=8$,
    to the nematically ordered state
for $p = .8$, starting from an empty lattice . Initially, the algorithm
produces a disordered state, which slowly relaxes to the almost
perfectly orientationally ordered state. The frames
a-f are taken after $ 1000, 4000, 8000, 12000, 20000$ and $36000$ MCS.
The asymptotic value of the order parameter is $Q=0.989$ and density,
$\rho = 0.758$.}
    \label{fig:fig2}
  \end{center}
\end{figure}

  We have studied this problem by Monte Carlo simulations using a
deposition-evaporation algorithm: At each time-step, we make an attempt to
deposit a rod with probability $p$, or evaporate one of the existing rods
with probability $1-p$. If it is a deposition attempt, we choose a
horizontal or vertical orientation with probability $1/2$ each, and pick
one of the sites of the lattice at random, and attempt to deposit the rod
there in the chosen direction. The attempt is successful if that site, and
the next $k-1$ sites in the chosen direction are all empty. Otherwise, the
attempted move is rejected. We use period boundary 
conditions in both the horizontal and vertical directions.
It is easy to see that this deposition-evaporation dynamics is ergodic and 
satisfies detailed balance, with the parameter $p$ related to activity 
$z$ by

\begin{equation}
z= \frac{\rho ~p}{ 2 k ( 1 - p) }.
\end{equation}

The deposition-evaporation dynamics does not conserve the number
of rods, $n$ and the number of vertical and horizontal rods   
$n_v$  and $n_h$ vary with time. 
In our Monte Carlo studies, we studied different values of $k$, from $2$ 
to $12$. We varied the deposition rate $p$
and monitored the density $\rho$
and the  order parameter $Q$.

Our main result is the evidence for existence of a orientationally ordered 
phase for intermediate densities of rods for $k \geq 7$.  The {\it 
existence} of the nematic phase is very clearly established in our 
simulations.  A typical run is shown in 
Fig.~\ref{fig:fig2}, where we have taken $k =8$, on a   $80 \times 80$ 
lattice, for $p= 0.8$, starting from all empty initial state.  
In the beginning, both horizontal and vertical deposition events are 
equally likely, and the density builds up to nearly equilibrium value 
$\simeq .75$, but 
the order parameter $Q$ remains nearly zero. By the time $1000$ MCS, the 
density has nearly stabilized to its equilibrium value, but the order 
parameter is nearly zero. With time, the size of locally ordered domains
grow, until one of the domains nearly fills the full lattice. For our 
lattice, this happens by around $16000$ MCS. At this stage, the order 
parameter is nearly $1$, and it shows only very small fluctuations away 
from this value for the full length of simulation ( about $10^7$ MCS). The 
average value of $Q$ in this run was $0.989$. Of course, the ordering is 
equally likely to be  horizontal or vertical. 

We have tested the results for different initial configurations, namely,
all empty, all vertical, all horizontal, half filled, etc. to ensure
sufficient time has been given for equilibration, and the averages 
calculated are independent of initial conditions. 

\begin{figure} \includegraphics[scale=.4, angle=270]{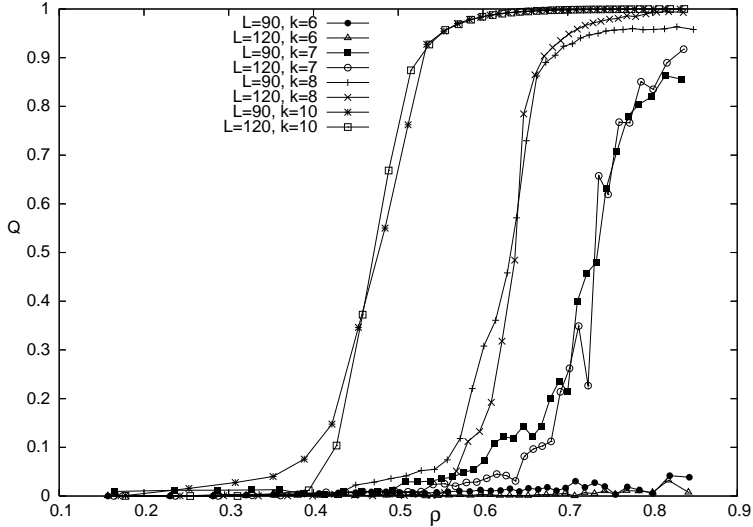} 
\caption{Order parameter $Q$ as a function of densities $\rho$
is shown for different $k$ and $L$.} 
\label{fig:fig3} \end{figure}

In Fig.~\ref{fig:fig3} we show that the order parameter $Q$ as a function
of density $\rho$ for different length of rods $k=6,7,8,10$ and two system
sizes $L=90$ and $120$. We did not find any isotropic-nematic transition
for short rods $k \leq 6$. However, as our simulations only cover the 
range $ 0 \leq \rho \leq .85$, we can only conclude that if $k_{min}$ the 
lowest value of $k$ that shows a nematic phase, then $ k_{min} \leq 
7$.

  For long rods $k \geq 7$, a  nematic phase is
observed for intermediate densities. The critical density for the onset of 
nematic order decreases with increasing $k$. 
This is consistent with the theoretical expectation that for very large 
$k$, the critical density should be related to that found in the problem of 
ultra-thin needles  in a continuum. If the critical density of needles of 
length $1$ in continuum is $A_c$, the critical density for the isotropic to 
nematic transition $\rho_{c1}$ should be approximately equal to $A_c/k$, 
for large $k$ [$A_c  \approx  6$, from simulations].    

\begin{figure} \includegraphics[scale=.4, angle=270]{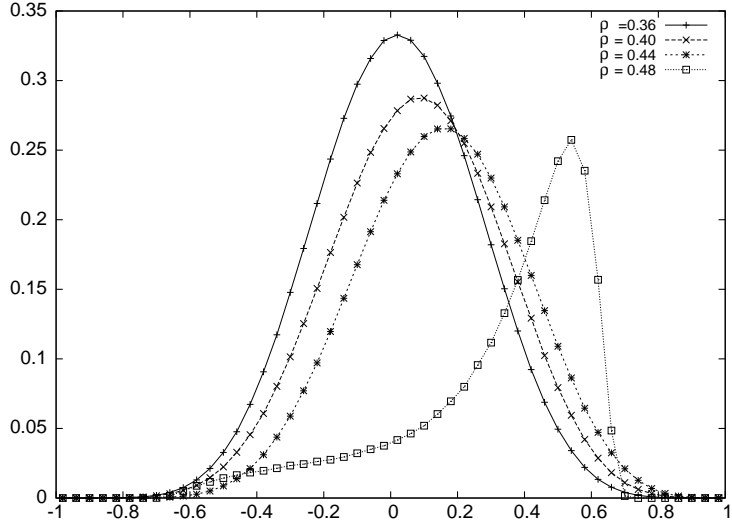} 
\caption{Normalized distribution of $n_v - n_h$ for $k=10$ and $L=120$
is shown for different values of densities $\rho$.} 
\label{fig:fig4} \end{figure}

The point of onset of the nematic order is not so easy to determine from 
the data.
In Fig.~\ref{fig:fig4} we show the distribution of $n_v - n_h$ 
for $k=10$ and $L=120$ for different densities $\rho$.
For $\rho < 0.36$ the distribution is centered around 0
implying that there is no preferred orientation in the configurations
and that the system is in isotropic phase.
While $\rho > 0.4$, the distribution shifts to the right
indicating onset of a nematic phase. 
Also, the order parameter 
increases gradually, indicating a continuous phase transition.
Since we are not able to determine the critical point with very good 
precision, we did not attempt to  study the critical exponents 
characterizing the transition.   As  there are two competing ordered 
states near the transition, one would expect this transition to be in the 
Ising universality class.

\section{The high density transition} \label{sec:high}

For intermediate densities, the state of the system is
well-approximated by the fully aligned state in which all the rods are
in the same orientation (say horizontal). We can determine the entropy
of the system approximately, by considering it as a perturbation from
the fully aligned state. The calculation of the entropy of the
fully aligned state density $\rho$ is straight-forward, and  reduces to 
the calculation of a
1-dimensional problem. The number of ways of putting ($\rho L/k$) rods
of length $k$ on a one dimensional line of $L$ sites is easily seen to
be
\begin{eqnarray}
\Omega_{nem} (L, \rho) = \frac{[L(1-\rho) +\frac{1}{k} \rho L
]!}{[L(1-\rho)]! (\frac{1}{k} \rho L)!}
\end{eqnarray}

>From this, it is easily seen that entropy per site in the thermodynamic
limit $L \rightarrow \infty$ in this reference nematic state is given by
\begin{eqnarray}
S_{nem}(\rho) = (1 - \rho + \frac{\rho}{k}) \log (1 - \rho + \frac{\rho}{k})
- (1 - \rho) \log (1 - \rho) - \frac{\rho}{k} \log \frac{\rho}{k}
\end{eqnarray}
For $\rho = 1 - \epsilon$, this tends to zero as $\epsilon$ tends to zero
as
\begin{eqnarray}
S_{nem}(\rho = 1 - \epsilon) = \epsilon \log \frac{1}{k \epsilon} -
\epsilon + {\rm higher~~order~~terms~~in~~} \epsilon
\end{eqnarray}

We can expand $\Omega(z_h,z_v)$ in powers of $z_v$, for a fixed value of
$z_h$. We treat $z_v$ as  small, and only at the end of
calculation put $z_v = z_h$
\begin{eqnarray}
\Omega(z_h,z_v) = \Omega(z_h,0) [ 1 +  z_v N  f_1 + z_v^2 ( N^2 f_1^2/2 + 
N 
f_2) + \cdots ].
\end{eqnarray}
where $N$ is the number of sites in the lattice. $f_1$ is expressible in 
terms of the probability of finding $k$
consecutive empty sites in the vertical direction in the reference
state. Since the configurations of different horizontal rods are
independent in the reference state, and the probability of finding a
random site unoccupied is $\epsilon$, we get $f_1 = \epsilon^k$. Using 
cumulant expansion, it is easy to 
see that 
\begin{equation}
S(\rho) - S_{nem}(\rho) =  z (1 - \rho)^k + z^2 f_2 + {\rm 
higher~~order~~terms}.
\end{equation}

The value of $f_1$ is quite small even for moderately large values of
$\epsilon$ and small $k$, Thus for $\epsilon = 1/4$ and $k=8$, the lowest
order term in the perturbation expansion of the fraction of vertical
$k$-mers in the nematic state is only of the order of $10^{-5}$. This is
much less than about $1\%$ fraction of vertical bonds seen in Fig. 2. Thus
the higher order terms are important.

 The second order term $f_2$ can be expressed in terms of the $2$-point
correlation functions of the unperturbed problem. The detailed form of
$f_2$ is somewhat complicated, and is omitted here.  It can be shown that
$f_2$ is of order $\epsilon ^{2 k} k^k$, which has the correct order of
magnitude.

However, for $\epsilon = 0$, the fully ordered state is not the most
likely state. The number of configurations of $k$-mers increases
exponentially with $N$ of the system as $\exp [ N S(\rho)]$,
where $S(\rho=1) > 0$.
It is straightforward to see that
\begin{eqnarray}
S(\rho=1) \geq \frac{\log 2}{k^2}
\end{eqnarray}
as there are two ways of covering a $k \times k$ square with straight
$k$-mers, and a lattice of size $kL \times kL$ can be broken up into $L^2$
such small squares.

\begin{figure} \includegraphics[scale=.32, angle=270]{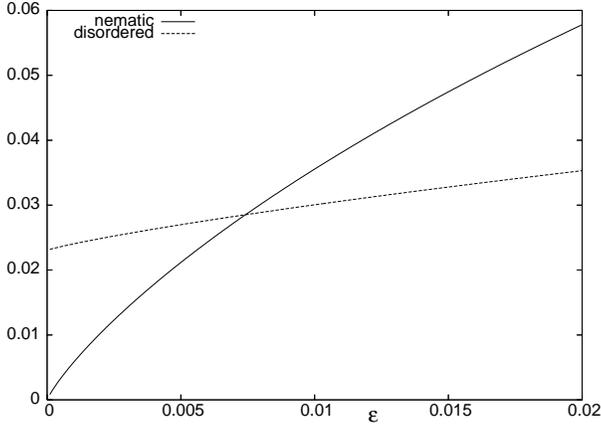} 
\caption{Entropy per site for nematic and disordered states as function
of $\epsilon$ for $k=8$.} 
\label{fig:fig5} \end{figure}

In fact we can get a better bound on $S(1)$ as follows: break the $kL
\times kL$ lattice in $L$ strips of width $k$ each. Now if $F_N$ is the
number of covering an $N \times k$ rectangle with $k$-mers, we easily see
that $F_N$ satisfy the recursion relation
\begin{eqnarray}
F_N = F_{N-1} + F_{N-k}.
\end{eqnarray}
This implies that $F_N$ increases as $\lambda^N$ where $\lambda$ is
the largest root of the equation
\begin{eqnarray}
\lambda^k = \lambda^{k-1} + 1.
\end{eqnarray}
Form this equation, it is easy to check that for large $k$,
\begin{eqnarray}
\lambda \approx  1 +\frac{1}{k} \log(\frac{k}{\log k}) + {\rm 
higher~~order~~terms~~~in~~}(1/k).
\end{eqnarray}
The entropy per site is then bounded from below by $(\log \lambda) / k$
and this gives the estimate $S(1) \approx \log k/ k^2$ for large $k$.

For densities away from the fully packed state an approximate state of
the system is obtained by first starting with a random configuration of
the fully packed lattice, and then remove a fraction of the rods at
random. This gives an approximate expression for the entropy of the
disordered state as 
\begin{eqnarray}
S_{dis}(\rho=1-\epsilon) \approx S_{dis}(\rho=1) + 
\frac{1}{k}[- \epsilon \log \epsilon - (1-\epsilon) \log (1-\epsilon)]
\end{eqnarray}

In Fig.~\ref{fig:fig5}, we have plotted $S_{dis}(\rho)$ and $S_{nem}(\rho)$
for $\rho$ near 1.  We note that for small $\epsilon$, 
$S_{nem}(\rho)$ increases with $\epsilon$ as $ \epsilon \log(1/ \epsilon)$, 
but $S_{dis}(\rho)$ has a weaker $\epsilon-$dependence, as it
varies as $(\epsilon/k) \log (1/\epsilon)$.  Clearly the
disordered state has higher entropy for $\rho=1$, but for lower densities, 
the
nematic state is favored. It is easy to verify that  the two
curves cross at $\rho_{c2} \approx 1 - C/k^2$,  for large k,  where $C$ is 
some constant.
Hence we expect a second phase transition from the nematic ordered state
to the disordered state as $\rho$ is increased beyond a density
$\rho_{c2}$.

Unfortunately, we are not able to study this second transition by our
Monte Carlo simulations. While the Monte Carlo algorithm is formally
ergodic for all $\rho \neq 1$, the relaxation times increases very fast as
the density increases and the system gets trapped in some small set of
nearby states. Our algorithm works well only for not too large values of
$\rho$ ( say $\rho < .85$).

We can only provide a qualitative description of this second phase
transition.  If we start with a fully packed configuration, and remove a
single $k$-mer, we get a state with $k$ unoccupied sites (`monomers').
These monomers occupy consecutive sites in the horizontal or vertical
direction. If the $k$-mers are allowed to diffuse, equivalently the 
monomers
can diffuse, but may remain  bound together as a molecular bound
state of $k$ monomers. However, at higher densities of such defects, 
these monomers
become unbound. In the nematic state, the monomers form a nearly ideal 
gas, and the larger entropy of the gas of monomers makes the nematic 
phase  preferred  thermodynamically.  Characterizing this phase transition, 
in particular 
to determine if it is first order or continuous is an interesting open
problem.

\acknowledgments{ We thank M. Barma, K. Damle, D. Das, M. D. Khandkar, J.
L. Jacobson and S. N. Majumdar for their comments on an earlier version of
this paper. This research has been supported in part by the Indo-French
Center for Advanced Research under the project number 3402-2.}

\end{document}